\documentclass[a4paper,11pt,notitlepage]{article}
\usepackage{graphicx}
\usepackage{amsmath}
\usepackage[a4paper]{geometry}
\usepackage{typehtml}
\usepackage{epsf}
\usepackage{amsfonts}
\usepackage{amssymb}

\begin{document}

\author{L. Vanel, E. Cl\'{e}ment\\Laboratoire des Milieux D\'{e}sordonn\'{e}s et H\'{e}t\'{e}rog\`{e}nes \\UMR 7603 - Bo\^{\i}te 86\\Universit\'{e} Pierre et Marie Curie,\\4, Place Jussieu, 75005\ Paris, France\\\bigskip\bigskip}
\title{Pressure screening and fluctuations at the bottom of a granular column}
\maketitle
\begin{abstract}
We report sets of precise and reproducible measurements on the static pressure
at the bottom of a granular column. We make a quantitative analysis of the
pressure saturation when the column height is increased. We evidence a great
sensitivity of the measurements with the global packing fraction and the
eventual presence of shear bands at the boundaries. We also show the limit of
the classical Janssen model and discuss these experimental results under the
scope of recently proposed theoretical frameworks.\newline pacs: 46.10+z,05.40.+j,83.70.Fn
\end{abstract}

\section{Introduction and background}

Mechanical description of static granular assemblies is an old but still open
problem which recently got revisited from different perspectives \cite{PDM98}.
In a classical and well established approach, practitioners of soil mechanics
developed a conceptual framework in the spirit of a plasticity
theory\cite{Schoefield68,Feda82}. This point of view based on an incremental
description of stress-strain relations is suited in practice for numerical
implementation. To close the description and introduce real granular effects
such as dilatancy, Mohr-Coulomb plasticity etc.. , many rheological parameters
have to be introduced on a semi-phenomenological basis\cite{Nedderman92}.
Hence, the constitutive laws usually proposed are often complex and rather
difficult to handle since they can be non-linear, non-differentiable and
anisotropic \cite{Harris92,Norris97}. As a consequence, the outcomes usually
based on numerical calculations and applied to complicated flow histories are
hard to follow and explain physically. Clearly, from a fundamental point of
view (maybe not on a practical point of view!), this approach is
unsatisfactory and at the moment, there is a vivid interest in establishing a
deeper understanding of the passage from a description of local granular
contacts and force distributions, to a macroscopic description of
stress-strain relations\cite{Cundall83,Goddard98}. A general feature observed
both in experiments \cite{Dantu57,Travers86,Tsoungui98,Delyon90,Liu95} and in
simulations \cite{Radjai96,Ouagenouni97,Radjai98,Eloy97} is the very
heterogenous and anisotropic character of the force network arising from the
intergranular contact geometry and the frictional properties of these forces.
The difficulty of the problem is such that no rigorous macroscopic theory of
granular mechanics is so far available. Moreover, how this disorder relates to
the dispersion and the fluctuations of macroscopic stress measurements is an
important issue that remains open \cite{Miller96,Ngadi98}. Recently, based on
phenomenology and properties of symmetry, theoretical works have suggested new
sets of closure relations for mechanical equations which aim to render the
existence of long range anisotropic structures in static assemblies, commonly
referred to as vaults. These relations contain a large set of possible
histories for the granular material\cite{Bouchaud95,Edwards96}. Besides, the
passage from a microscopic to a macroscopic description was undertaken using
various models of force networks, the resolution of which is based on cellular
automatons \cite{Coppersmith96,Claudin97,Hemmingsson97}, stochastic
toy-models\cite{Eloy97,Socolar98} and analytical solutions of stochastic
stress transport equations \cite{Claudin98a}.

The case of a container filled with grains is a basic situation where the
presence of boundaries is a natural initiator of a pressure screening effect
that Janssen described\ last century in a pioneering contribution
\cite{Janssen}. Quite surprisingly, the simple Janssen analysis is still used
as a basis for the design of silos \cite{Schwab94,Thompson96} despite many
difficulties in testing any models which come from a lack of reproducible
experiments. Large fluctuations of the mean vertical pressure are generically
observed in silos \cite{Brown70}. Data shows large differences for tests in
identical conditions \cite{Shaxby23} and a strong dependency on the filling
procedures \cite{Jotaki77}. The numerous theoretical refinements proposed to
improve the Janssen analysis (see for example Ref. \cite{Nedderman92}) fail to
take into account the existence of widely scattered pressure measurements and,
in our opinion, this is a reason why an attempt to comparatively differentiate
all theoretical predictions is most often doomed to failure \cite{Lenczner63}.

In this work, we design an experimental set-up with the goal of precisely and
reproducibly measuring the mean pressure as well as its fluctuations at the
bottom of a granular column. We present two sets of well defined procedures
which allow to obtain reproducible measurements. We quantitatively analyze
these results in the form of an adaptation of the Janssen's classical model,
and we discuss several theoretical models that could be used to compare our
results with.

\section{Experimental set up}

\subsection{Apparatus description}

A sketch of the experimental set-up is shown on Fig.1. A mass $M$ of grains is
poured into a vertical cylinder of inner radius $R=20mm\ $and lies on the
horizontal top surface of a piston. The piston is designed in such a way that
it never touches the cylinder walls. Thereby, the average pressure of the
granular material on the piston is entirely transmitted to an electronic
scale. The scale is a horizontal beam of high effective stiffness
$K=2.10^{4}N/m$, the deformation of which is detected with a strain gauge
bridge which has the property to eliminate temperature drift effects in strain
measurements. The corresponding force is measured in unit of mass with a
precision $\Delta m=0.1g$ and it will be referred to as the \textit{apparent
mass} $M_{a}$, since its value is different from the total mass $M$ of the
granular medium. The system ''beam+piston'' rests on a mechanical elevator
whose vertical displacement $\Delta Z$ can be varied and measured with a
micrometric screw.

\subsection{Filling procedure}

We used this apparatus to measure the apparent masses of various species of
grains such as monodisperse glass spheres, polydisperse iron shots or rugged
quartz grains. All the results presented in this paper concern monodisperse
glass spheres of diameter $d=2mm$. In order to vary the initial packing
fraction, two filling methods are used (see sketch on Fig.1). In method $1,$
the grains fall from a hopper located on the top of the column and due to the
ramming effect of falling grains a rather compact packing is obtained. In
method $2,$ the grains fill first an intermediate inner cylinder that
initially rests on the piston. Next, the inner cylinder is slowly removed so
that the grains gently flow into the outer cylinder and settle rather loosely.

\subsection{Volume fraction measurements}

Throughout the experiments, the average volume fraction of the grains was
estimated by monitoring the total height $H$ of the granular column. The top
level of filling is detected using a digital vernier calliper (precision
$0.01mm$). We checked that within a precision of one bead radius, the top
surface was almost flat. This is probably a consequence of the rather small
ratio between the column radius and the bead size : $R/d=10$ which prevents
the development of a well defined slope at the free surface$.$ A thin and flat
cardboard disc of mass $m=0.3g$ was used to obtain a well-defined top surface
and reduce perturbations from the rod when coming in contact with the granular
material. The bottom displacement of the column is known from the elevator's
micrometric screw. The average granular density is therefore computed as
$\rho=M/\pi R^{2}H,$ and the packing fraction is then $\nu$ $=\rho/\rho_{b}$,
where $\rho_{b}$ is the grain density. The height measurement is always done
after the corresponding apparent mass measurement since the latter is very
sensitive to all kind of perturbations as we will see in the following.

\section{Obtention of reproducible pressure measurements}

\subsection{Measuring stresses in granular assemblies}

There is a fundamental and technical difficulty in measuring stresses in
granular assemblies. One physical reason is rooted in the hysteretic character
of the local friction forces between the grains and with the system boundaries
which prevent the system from returning to the initial equilibrium after a
perturbation. We stress that it is not necessary to reach plastic deformation
on a large scale in order to observe an irreversible change in the equilibrium
state. For instance, a simple spring-mass system with friction displays
hysteretic properties \cite{Duran98} and the coupled frictional and elastic
contacts between grains are expected to display similar features
\cite{Mindlin53}. Since every pressure probe is by construction associated
with some displacement of a reference surface like a membrane or a piston, the
measuring process may modify in return the force distribution in an
irreversible way and therefore, the measurement values will depend on the
specific perturbation history of both the probe and the material. One remedy
could be in a quest for the least disturbing probe like a very stiff spring or
a feedback device where position is controlled, but this might create other
problems such as pathological coupling with temperature (due to thermal
dilation\cite{Clement97}), a decrease in the sensitivity and/or incontrollable
instabilities. This is exactly the kind of problem we address in this article.
We consider the granular column and the pressure probe as a paradigmatic
situation where all the difficult aspects of solid friction, boundaries and
pressure measurement are \textit{a priori} coupled in a complex way. The basic
idea of the measurement device we propose, and in particular the use of a
mechanical elevator, is to produce a statistical distribution of the many
metastable states corresponding to a specific average equilibrium with the
piston. We want to know, for example, whether the dispersion and
non-reproducibility of results currently reported in the literature is due to
an intrinsically wide distribution of metastable states or if other ''hidden
parameters'' should be considered to rationalize the results. In the
following, we report two distinct ways of producing controlled and
reproducible measurements, and furthermore, we describe basic observations
that could explain why previous experiments have yielded non-reproducible data sets.

\subsection{ Measurement procedures}

Two different kinds of experimental procedures are investigated. Both lead to
a set of reproducible results within a rather small fluctuation scale. They
are both based on a slow downward displacement of the system ''beam+piston''
at a slow velocity (typically $20\mu m/s$). Note that importantly, the
granular column seems to slide as a whole since we realize that any downward
displacement of the bottom piston is also evidenced as a simultaneous downward
displacement of the column's top. This procedure has \textit{a-priori} two
remarkable advantages: (i) the kinetic energy of the falling grains during the
filling stage which is partially stored as elastic energy of the beam, is
relaxed, and therefore there is a memory loss of the initial pouring dynamics;
(ii) the downward motion of the piston allows the granular column to slide
down such that the friction forces at the walls should be fully mobilized and
directed upwards. This is crucial for a quantitative analysis since generally,
the theoretical models assume that the granular material is at the sliding
limit at the walls. This important step was missing in previous experimental
reports that we know. Generally speaking, the system would reach an
equilibrium position depending specifically on the kinetic energy due to the
pouring procedure, the effective distribution of friction forces at the
boundaries, and the actual displacement of the pressure probe (related to the
spring stiffness).

Note that in a preliminary set of experiments\cite{Clement97}, we reported a
large coupling of the pressure with small temperature variations. We linked
this effect to problems of differential thermal dilation between different
parts of the setup. Therefore, a double isothermal container was used to
reduce the drastic effect of temperature on the plexiglass column as much as
possible (this material has a large dilation coefficient and is a poor
temperature conductor). In the new set-up, problems of temperature are
dampened using a material with a very low temperature dilation coefficient
such as an iron-nickel alloy. It turns out that this precaution is sufficient
to obtain measurements independent of the actual temperature values. In fact
we observe that after each downward displacement, there is a time gap of
typically $30s$ which is large enough to relax our mechanical system and to
make a measurement, but which is small enough to be able to neglect the
effects of temperature drifts.

\subsubsection{Descent experiments}

The first kind of experiment probes the effect of accumulated series of
downward displacements of the piston. It will be referred to as the
\textit{descent experiment.} Note that in a preliminary report, the procedure
has been presented as well as some of the effects we discuss in the following
\cite{Vanel98}. The vertical displacements are performed with a fixed
amplitude of $0.125mm$ ($1/16$\ of bead size) on a total distance of $20mm$
corresponding to one column radius. The displacement amplitude has been chosen
for practical convenience, but we checked that changing its magnitude does not
modify the reported behavior. Fig. 2a shows a typical evolution of the
apparent mass both during the displacement of the piston and after the piston
was stopped. We observe that when the piston is slowly moved down, the
apparent mass abruptly decreases. The accelerated motion communicated to the
system piston+scale during this process is responsible for a slight
decompression of the scale spring which leads to this abrupt decrease. As soon
as the elevator stops, the apparent mass suddenly increases up to a point
where a slow relaxation of the apparent mass occurs and a stable value is
finally reached. Since a decrease in the apparent mass implies an upward
motion of the scale spring, it could be concluded that after the slow
relaxation the friction forces are not fully mobilized at the
walls.\ Nevertheless, as a practical statement, we note that the amplitude of
the relaxation effect on the apparent mass values is less than $2\%$ so that
the friction would be in any cases almost fully mobilized. Moreover, a
decrease of the apparent mass means that the overall screening effect becomes
stronger. This is in contradiction with a loss of friction mobilization and
leads us to think that the small upward displacement of the piston is not
really disturbing. One should also keep in mind that this displacement is less
than $0.5\mu$m, which is smaller than the typical distance to mobilize
friction \cite{Berthoud98}. In fact, the increase of the screening effect
during this slow relaxation may correspond to a slow reorganization of the
packing structure, but this point deserves a more thorough study. In this
report, we only concentrate on the final values of the apparent mass as shown
on Fig. 2a. Fig. 3 shows typical plots of the apparent mass as a function of
the piston displacement for a filling mass $M=300g$. This filling corresponds
to a height : $H\simeq8R$. Whatever filling method is used, there is a global
increase of the apparent mass with the number of displacement steps and,
eventually, it seems a steady state is reached. Around this global behavior,
fluctuations of a smaller scale are evidenced. The average packing fraction
changes with the number of displacement steps and also tends towards a steady
state. When the initial packing is rather compact ($\nu\simeq0.63$), a global
decompaction occurs, while there is a global compaction when the initial
packing is rather loose ($\nu\simeq0.58$). Furthermore, whatever the initial
packing fraction, the final packing fractions tend to be the same. The
reproducibility of the method can be evidenced on Fig. 3a where the results of
two independent experiments are superposed.

\subsubsection{Tapping experiments}

The second kind of experiment probes the effect of changing the granular
density with the use of vibrations produced by series of taps on the container
walls. It will be referred to as the \textit{tapping experiments}. Starting
from a loose piling, previous studies have shown that series of taps induce a
slow compaction effect\cite{Knight95}. Note that in this last reference,
tapping was a vertical impulse; here, we use series of side impacts. The
typical evolution of the apparent mass is shown on Fig. 2b. After a tap, there
is a sudden increase of the apparent mass. The main reason for this increase
is due to an hysteresis effect of the measurement device. A tap on the wall
can either break the frictional contacts between the grains and the walls or
make the piston vibrate. In both cases this should result in an additional
downward compression of the piston. After a tap, the piston cannot decompress
and return to its initial equilibrium because of the granular column
frictional resistance to motion. In order to release the dynamical compression
effect due to a tap, the piston is then slowly moved down and as a
consequence, there is a fast decrease of the load to a level which is quite
reproducible. A remarkable fact is that the final apparent mass (after a small
downward displacement) is independent of the impact itself but is very well
correlated to the average volume fraction, and this for the many independent
experiments that we did. On Fig. 2b, we also display the mass variation
obtained from the initial pouring procedure.\ The corresponding data points
are seen on this figure for a time less than $50s$. We observe that the
apparent mass indeed saturates at a value of about $80g$. Nevertheless, we
tested that this value is difficult to reproduce from one filling to the
other. On the other hand, after the first small and slow displacement $\Delta
z$, we obtain a value $50\%$ smaller which is reproducible from one experiment
to the other within a $5\%$ error bar! On Fig. 4, we display a typical plot of
the apparent mass $M_{a}$ as a function of the packing fraction for a filling
mass $M=300g$. Two independent sets of data are shown to illustrate the
reproducibility of the variation as well as the level of fluctuations. The
tapping procedure leads to a progressive compaction of the granular column
while the apparent mass is decreasing.

\section{Analysis of the Experimental results}

\subsection{A phenomenological description and a qualitative interpretation}

The results obtained in both descent and tapping experiments can be summarized
on a unique diagram. Fig. 5a shows such a diagram where the apparent mass is
plotted as a function of the packing fraction for a filling mass $M=300g.$ The
arrows indicate the direction of evolution in the course of each experiment. A
similar qualitative diagram is obtained with many other granular materials
like polydisperse steel spheres or angular quartz grains, but we only report
the results for $d=2mm$ glass spheres here. There is clearly a drastic change
in the static equilibrium with the experimental procedure followed. Data for
the\textit{\ tapping experiments} shows that a density increase of about $8\%$
induces a pressure decrease of about $20\%$ (Curve \{1\}). On the other hand,
the descent experiments show that density variations of less than $5\%$ induce
a pressure increase as large as $50\%$ (Curves $\{2\}$ and $\{3\}$). Moreover,
this last procedure shows a compaction effect for initially loose packings and
a decompaction effect for initially dense packing. Experiments on a 2D
Schneebeli medium show than in a descent experiment like ours, friction at the
wall produces shearing bands \cite{Pouliquen96}. In these experiments, the
assembly was initially rather compact and due to shearing, a global
decompaction occurred creating a radial density profile with a looser density
on the edges and a larger density at the center. We suppose that in our
descent experiments, a similar shearing effect occurs. Moreover, our results
are consistent with the classical vision of soil mechanics where many standard
tests have shown that shearing a dense medium produces a
decompaction\cite{Reynolds} and shearing a loose medium produces a
compaction\cite{Schoefield68}; this is what happens in curves $\{2\}$ and
$\{3\}$. In the limit of large deformation a so-called ''critical density'',
independent of the initial stage, is reached. This is probably what could
happen around $\nu o\simeq0.593$ where both curves cross. In our case, it
simply indicates that an identical density profile might be finally reached
from both sides. On the other hand, what happens in $\left\{  1\right\}  $ is
\textit{a priori} less clear. The way we understand these results is that the
vibrations produce a compaction effect which ''kills'' the density gradients
developed through the shear bands. After a \ tap, the displacement of the
piston is small enough to avoid the formation of shear bands, and even if a
shear band was initiated, the next vibrational shock is likely to destroy it.
As a consequence, we have a more homogeneous granular packing along the radius
and likely in the whole column. This interpretation is consistent with the
fact that the end of curve $\left\{  1\right\}  $, corresponding to a sequence
of tapping and descents, seems to join the beginning of curve $\left\{
3\right\}  $ which represents an early situation when the shear bands are not
yet developed. In the tapping experiment, we observe a decrease of the
apparent mass when the density is increased, which witnesses a stronger
screening effect of the boundaries. Such a dependence between friction and
density is reminiscent of recent experimental results \cite{Horwarth96} in
which the extraction force of a rod buried in sand is shown to increase
drastically with the packing fraction. At this point, to go beyond these many
conjectures, more information on the local density spatial distribution would
be needed. We are in the process of measuring this experimentally using an
invasive technique.

\subsection{Quantitative study of the pressure saturation}

\subsubsection{Comparison with the classical Janssen model}

In the following, we study the apparent mass as a function of the total mass
poured in the cylinder. As a guide-line for describing the results, we present
at first the Janssen classical analysis\cite{Janssen}. Last century, Janssen
proposed a simple heuristic argument to account for pressure distribution
$P(z)$ at the edges of a container filled with a granular material. The
argument is based on mechanical equilibrium of a horizontal slice of thickness
$dz$ and horizontal surface area $S=\pi R^{2}.$ The slice is submitted to the
action of its own weight, a pressure gradient from the surrounding granular
material and friction forces $dF_{frict}$ from the lateral walls such that:%

\begin{equation}
-\nabla_{z}P(z).Sdz+\rho gSdz-dF_{frict}=0\tag{1}\label{equ1}%
\end{equation}

where $\rho$ is the granular density. The core of the model assumes that the
vertical pressure is transferred into a horizontal pressure via a constitutive
Janssen coefficient $K$ and that the friction forces are \textit{fully
mobilized} in the \textit{upward} direction. Thus the friction force acting on
a surface element $dS=2\pi Rdz$ is: $dF_{frict}=\mu KP(z)2\pi Rdz,$ where
$\mu$ is the solid friction coefficient of the grains with the wall. A rough
measurement of the coefficient of friction between glass beads and the wall
gives $\mu=0.4(\pm0.02)$. Integration of equation(\ref{equ1}) with the
boundary condition $P(0)=0$ yields an exponential saturation of the pressure
with depth $z$: $P(z)=\rho g\lambda(1-\exp(-z/\lambda))$. A central parameter
in the theory is the characteristic length $\lambda$ which accounts for a
pressure screening effect due to the boundaries: $\lambda=R/2\mu K$. In fact,
the theory assumes that the stress distribution is uniform across any
horizontal section of the material. This assumption turns out to be false and
many theoretical refinements have aimed to correct it (see, for example, ref.
\cite{Nedderman92}). However, the predicted pressure saturation curves are
qualitatively very similar to each other. In the following, we only consider
the simple saturation equation predicted by Janssen, except that it is
expressed in unit of mass. The saturation mass, $M_{\infty}=\rho\pi
R^{2}\lambda$, the mass $M$ of grains poured inside the column and the
measured apparent mass $M_{a}$ should agree with the equation :%

\begin{equation}
M_{a}=M_{\infty}(1-\exp(-M/M_{\infty}))\tag{2}\label{equ2}%
\end{equation}

We experimentally tested this law by varying the mass of grains in the silo.
To change the filling mass, the silo was entirely emptied and filled again
with the desired amount of grains. Hence, data obtained for different filling
mass corresponds to \textit{completely independent sets of experiments}. Since
we have good reasons to believe that the packings obtained are more
homogeneous, we now focus on the data obtained with the second measurement
procedure, i.e. the \textit{tapping} \textit{experiments}. On Fig. 6, the
apparent mass is plotted as a function of the filling mass for an average
packing fraction $\nu=0.585\pm0.005$. Note that the straight line on the
picture would represent a perfectly hydrostatic behavior. We clearly observe a
saturation of the bottom pressure as the most elementary Janssen vision would
predict. On the other hand, at a quantitative level, we evidence that the
experimental data points are systematically \textit{above} Janssen equation
(\ref{equ2}) when the fitting parameter $M_{\infty}$ takes the value of the
experimental saturation mass (see dashed curve on Fig.6 obtained with:
$M_{\infty}=53g$). This remains true for every saturation curve we obtained
with other volume fractions.

\subsubsection{Analysis using a modified Janssen model}

To account for the experimental fact that the hydrostatic behavior is more
pronounced than Janssen prediction, we propose a two-parameter model where is
explicitly introduced a finite-size zone the apparent behavior of which is
\textit{purely hydrostatic}. At the most elementary level, the hydrostatic
zone is viewed as an horizontal slice of mass $M_{0}$ located at the bottom of
the column, while the rest of the column is supposed to behave according to
Janssen's differential equation (\ref{equ1}). The resulting equations are :%

\begin{equation}%
\begin{array}
[c]{l}%
-\text{for }M\leq M_{0},M_{a}=M,\\
-\text{for }M>M_{0},M_{a}=M_{0}+M_{\infty}^{C}\left[  1-\exp\left(
-(M-M_{0})/M_{\infty}^{C}\right)  \right]
\end{array}
\tag{3}\label{equ3}%
\end{equation}

The thin solid curve on Fig. 6 is a best fit of the experimental data points
with the previous equation (\ref{equ3}). The corresponding values of the
fitting parameters are $M_{0}=13.4g$ et $M_{\infty}^{C}=39.6g$. Although the
two-parameter model is based on a very crude assumption, the fit is quite good
and clearly captures the main features of the experimental results. From the
data obtained with the many \textit{tapping experiments} we performed, we
measured the corresponding sets of parameters: $M_{0}$ and $M_{\infty}^{C}$,
for other values of the packing fraction. The dependency of the fitting
parameters with the packing fraction is shown on Fig. 7. We find that $M_{0}$
has only a slight tendency to decrease with $\nu$ while the parameter
$M_{\infty}^{C}$ is clearly a decreasing function of the packing fraction.
From the values of $M_{\infty}^{C}$ and the relationship $M_{\infty}^{C}%
=\rho\pi R^{3}/2\mu K_{eff}^{C}$, an effective Janssen coefficient
$K_{eff}^{C}$ is determined with the hypothesis that the coefficient of
friction is independent of the packing fraction. As a comparison, we also
determined the effective Janssen coefficient $K_{eff}$ obtained with the
original Janssen model where the saturation mass $M_{\infty}$ is simply the
sum of $M_{\infty}^{C}$ and $M_{0}.$ In both cases, the effective Janssen
coefficients increase with $\nu$ (see Fig. 8).

\section{Discussion}

In the previous report we experimentally determined the shapes of the
saturation curves for columns in static equilibrium with packings of different
densities. We evidenced that the effective form of the saturation curve is
strongly dependent on the density and as a rough statement, we find (i) that
the effective screening length has a tendency to decrease when the density
increases, and (ii) that the saturation curves stay very close to the
hydrostatic curve for small fillings. We are aware that the use of a
two-parameter model to fit the data is not really satisfactory. Nevertheless
we believe that this investigation can be useful since any interpretative
theoretical model should be consistent with these experimental facts and
moreover, any theoretical saturation curve analyzed with our two-parameter
model should provide relations identical to the one obtained on Fig. 8.
However, from a classical perspective, we found \textit{a priori} surprising
that $K_{eff}$ stays smaller than unity while $K_{eff}^{C}$ is almost always
larger than unity. In the engineering literature\cite{Nedderman92}, it would
be said that the granular material is in an active state when $K<1$ and in a
passive state when $K>1.$ The real physical meaning of this classification is
usually related to the mobilization of friction forces \textit{inside the
granular material} and, \textit{a priori}, it is not obvious to which case our
experiments should correspond to.

Recently a theoretical prediction was made on the existence of a pronounced
hydrostatic zone for a column filled with grains of finite elasticity
\cite{deGennes98}. According to this theory, there is a zone at the top and at
the bottom of the column where the grain displacements relative to the walls
are much smaller than the minimum distance necessary to mobilize friction.
Here we are under the impression that our descent procedure should prevent
such a condition to occur. Another physical explanation for the hydrostatic
zone could be the following. Let us assume that the stresses actually
propagate along specific paths. The qualitative argument is that the weight of
grains close to the piston can be carried along stress paths that will never
reach the walls but instead, directly hit the piston. Then some of the mass
cannot be screened by the walls so that the real saturation curve should be
closer to the hydrostatic curve as compared to a simple Janssen law. Several
models actually predicts that the stresses propagate along specific
directions. This is the case of a model Bouchaud et al \cite{Bouchaud95} have
proposed recently, but it is also a consequence of the classical I. F. E.
hypothesis (Incipient Failure Everywhere). In addition, Socolar has found
using a stochastic toy-model that the average vertical force in a silo is
closer to the hydrostatic curve than the Janssen prediction. The curve
predicted by Socolar in Fig. 4 of ref. \cite{Socolar98} and our experimental
curve in Fig. 6 are very similar. Socolar also observed the emergence of long
stress chains which are reminiscent of experimental observations
\cite{Dantu57}. Those stress chains represent stress paths along which large
values of stress are carried. Thus, the qualitative interpretation we
described above seem to be supported by Socolar's results. The question arises
whether other theories which don't predict the existence of stress chains can
also describe the pronounced hydrostatic behavior we observed experimentally.
A precise quantitative comparison between various theoretical models and our
experimental data is the aim of a subsequent paper \cite{Claudin98b}.

\section{Conclusion}

In this paper we investigate the fluctuations and variations of the average
pressure at the bottom of a granular column. First we design an apparatus and
two experimental procedures which allow to obtain reproducible measurements.
For both procedures, we are able to separate a level of fluctuations and a
level of systematic variation which are very well correlated with the average
packing fraction. One of these operating modes is thought to provide
information on rather homogeneous packings and we perform a quantitative
analysis of the pressure saturation curves for different packing fractions. We
find that the data points are systematically above the simplest prediction one
can make using a Janssen model. We make a simple extension of this model and
introduce an effective hydrostatic zone at the bottom of the column as a new
fitting parameter. The agreement with the experimental results is quite good
and we find that the Janssen constant which is the second fitting parameter,
shows a systematic increase with the packing fraction. On the other hand, the
bottom hydrostatic mass shows a weak variation within the experimental errors.

The second experimental procedure is thought to achieve a situation where the
downward motion of the grains and the friction with the boundaries produce
localized shear bands. Therefore a systematic variation of the density is
expected along the column radius, i.e. the packing fraction at the walls is
less than at the center. Although this density variation is very small, the
effect on the mechanical equilibrium is drastic since the apparent saturation
mass shows a systematic increase and reaches values which are twice as large
as the one obtained with an homogeneous column. We are aware that this series
of experiments and the interpretation we propose, though encouraging, are
preliminary. The fact that we find reproducible data sets calls for more
experimental work with other materials, other boundary conditions and larger
column sizes. Moreover, we need a systematic series of tests in order to
discriminate clearly between all different theoretical approaches and possibly
reach a well established vision of this elementary but still unraveled problem
of the static equilibrium of a granular assembly in a column\cite{Claudin98b}.

\clearpage%

\begin{figure}
[ptb]
\begin{center}
\includegraphics[
height=7.8156cm,
width=12.7536cm
]%
{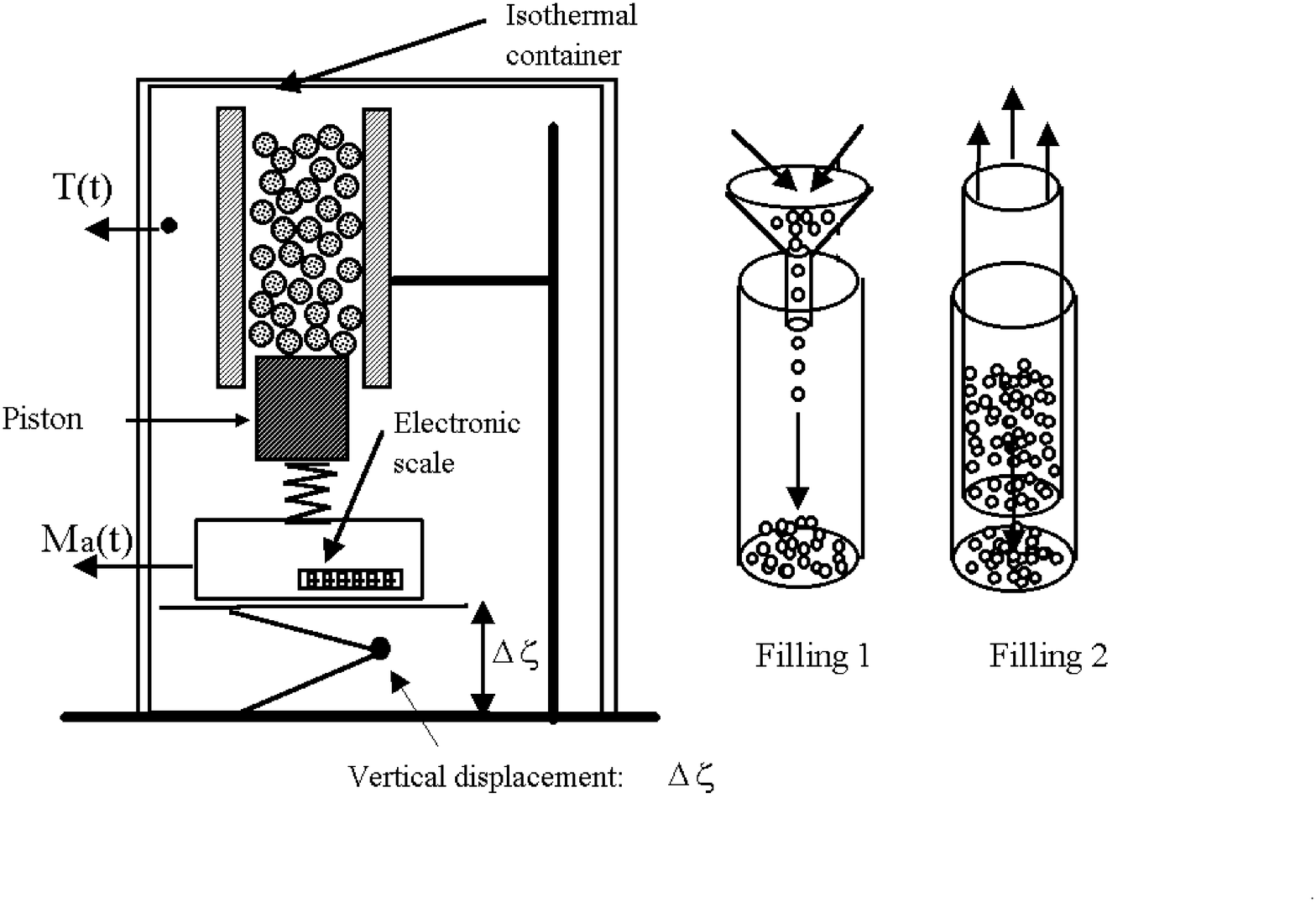}%
\caption{Left picture: sketch of the experimental apparatus. Right picture:
sketch of the two filling methods where the arrows indicate the motion of the
grains and/or the motion of the intermediate inner cylinder.}%
\end{center}
\end{figure}

\begin{figure}
[ptb]
\begin{center}
\includegraphics[
height=15.6378cm,
width=12.1715cm
]%
{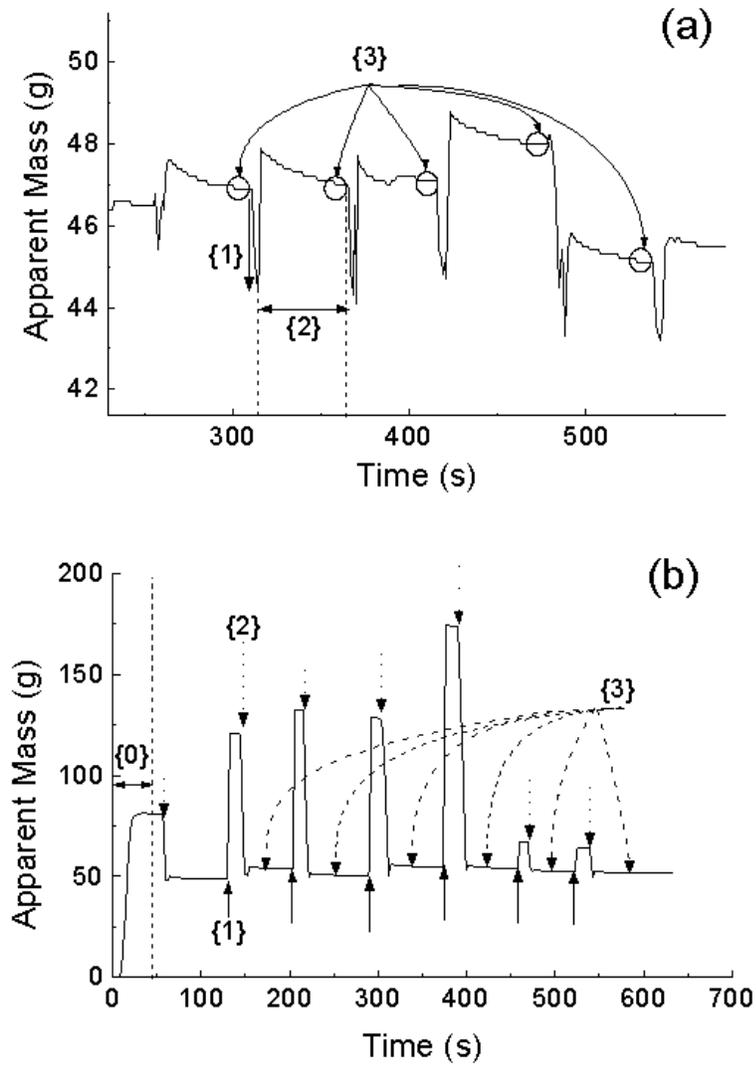}%
\caption{Evolution of the apparent mass $M_{a}$ with time: (a) during a
\textit{descent experiment}: \{1\} indicates a descent of the piston; \{2\}
indicates a period of rest for the piston; \{3\} indicates the measured values
of the apparent mass. (b) during a \textit{tapping experiment:} \{1\}
indicates a tap; \{2\} indicates a descent of the piston; \{3] indicates the
measured values of the apparent mass. The \{0\} sign indicate the filling
stage. The discrete steps observed correspond to the finite resolution of the
measuring scale.}%
\end{center}
\end{figure}

\begin{figure}
[ptb]
\begin{center}
\includegraphics[
height=16.8086cm,
width=12.0463cm
]%
{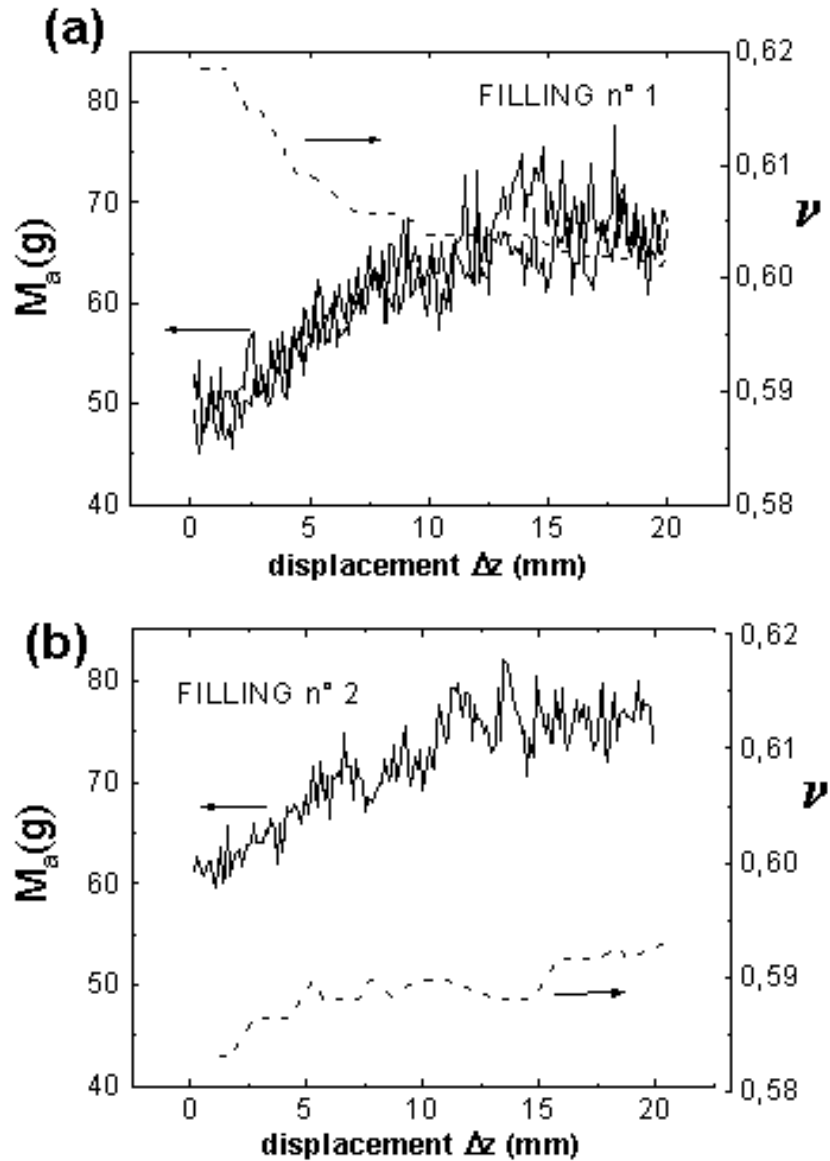}%
\caption{\textit{Descent experiment}: Apparent mass $M_{a}$ (solid line) and
packing fraction $\nu$ (dashed line) as a function of the vertical downward
displacement $\Delta Z$ for a filling mass $M=300g$. (a) filling method $1$,
the thin and the thick solid lines are two independent experiments; (b)
filling method $2$.}%
\end{center}
\end{figure}

\begin{figure}
[ptb]
\begin{center}
\includegraphics[
height=9.0896cm,
width=11.3346cm
]%
{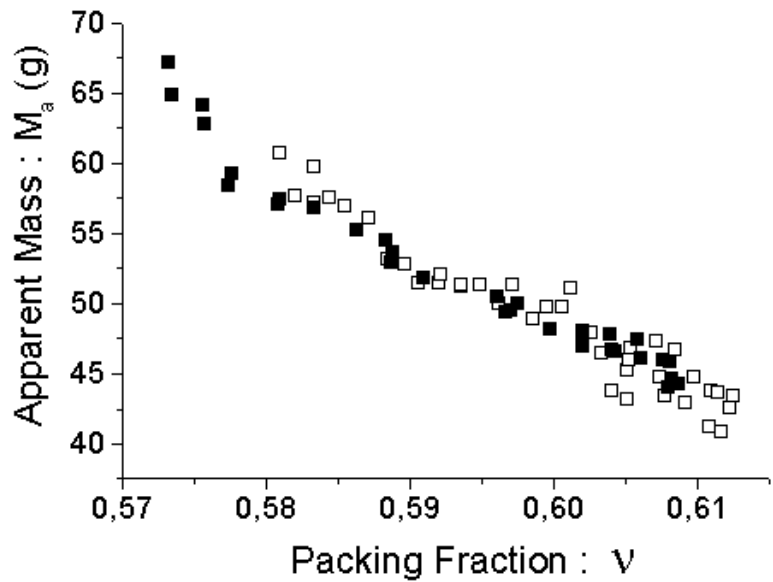}%
\caption{\textit{Tapping experiment: }Apparent mass $M_{a}$ as a function of
the packing fraction $\nu$ for a filling mass $M=300g.$ Open and closed
squares refer to two distinct experiments.}%
\end{center}
\end{figure}

\begin{figure}
[ptb]
\begin{center}
\includegraphics[
height=9.3027cm,
width=12.11cm
]%
{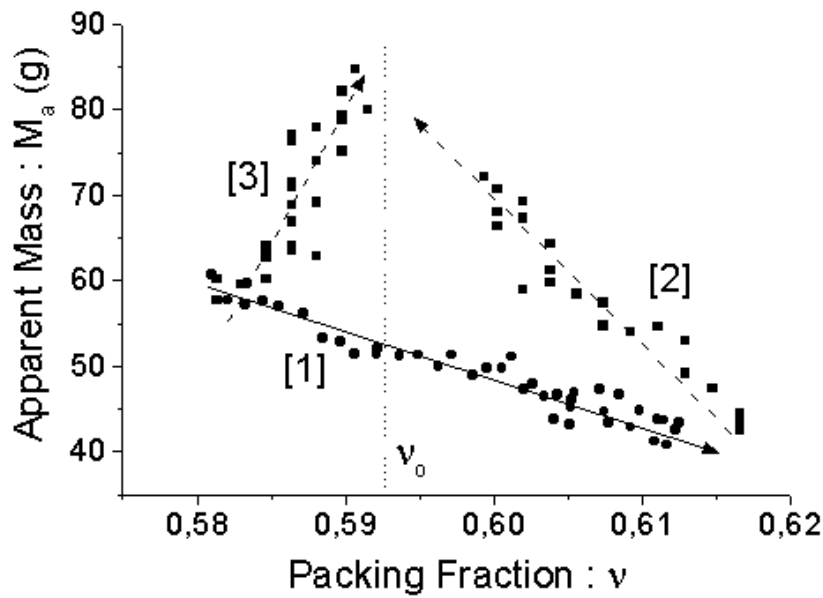}%
\caption{Apparent mass-Packing fraction diagram with a filling mass $M=300g$
for monodisperse glass spheres. Sign [1] indicates a tapping experiment, sign
[2] a descent experiment with filling method n${{}^{\circ}}1,$ sign [3] a
descent experiment with filling method n${{}^{\circ}}2.$ Arrows indicate the
evolution in the course of each experiment.}%
\end{center}
\end{figure}

\begin{figure}
[ptb]
\begin{center}
\includegraphics[
height=9.4389cm,
width=12.0375cm
]%
{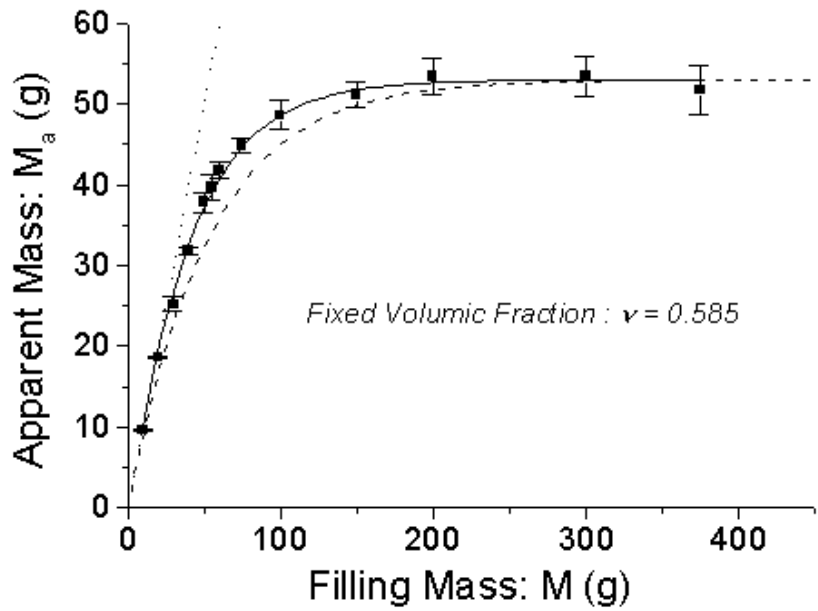}%
\caption{\textit{Tapping experiments: }Apparent mass $M_{a}$ as a function of
the filling mass $M$ for a packing fraction $\nu=0.585\pm0.005.$ The straight
line indicates an hydrostatic behavior; the dashed curve is a fit with Janssen
prediction (equation \ref{equ2}) where $M_{\infty}=53g$; the thin solid curve
is a fit with the two-parameter model (equation \ref{equ3}) where
$M_{0}=13.4g$ and $M_{\infty}^{C}=39.6g.$}%
\end{center}
\end{figure}

\begin{figure}
[ptb]
\begin{center}
\includegraphics[
height=19.2402cm,
width=12.0133cm
]%
{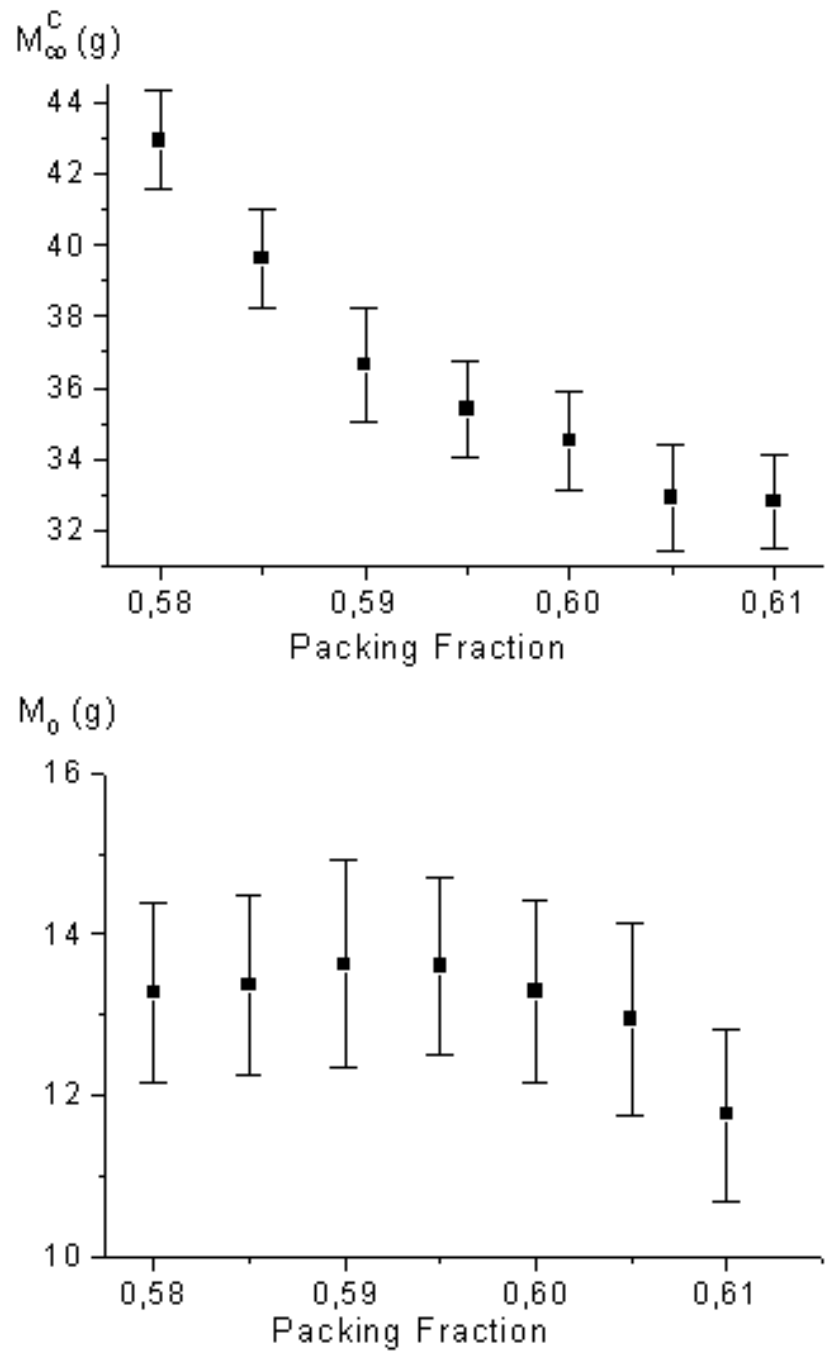}%
\caption{\textit{Tapping experiments: }Dependency of the two parameters
$M_{0}$ and $M_{\infty}^{C}$ with the packing fraction.}%
\end{center}
\end{figure}

\begin{figure}
[ptb]
\begin{center}
\includegraphics[
height=9.0193cm,
width=11.965cm
]%
{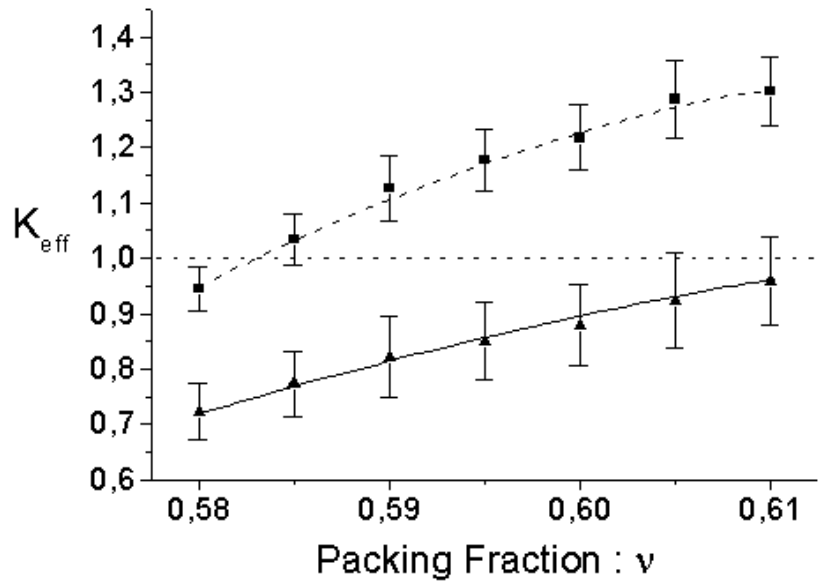}%
\caption{\textit{Tapping experiments: }The effective Janssen coefficient
$K_{eff}$ as a function of the packing fraction: ($\blacktriangle)$ $K_{eff}$
estimated from the Janssen saturation mass $M_{\infty};$ ($\blacksquare)$
$K_{eff}$ estimated from the corrected saturation mass of the two-parameter
model $M_{\infty}^{C}.$}%
\end{center}
\end{figure}
\end{document}